\documentclass[letter]{aa} % for the letters 
\usepackage{graphicx}
\usepackage{txfonts}
\begin{document}

   \title{KIC 9821622: An interesting lithium-rich giant in the \textit{Kepler} field\thanks{Based on observations obtained at the Gemini Observatory, which is operated by the Association of Universities for Research in Astronomy, Inc., under a cooperative agreement with the NSF on behalf of the Gemini partnership: the National Science Foundation (United States), the National Research Council (Canada), CONICYT (Chile), the Australian Research Council (Australia), Minist\'{e}rio da Ci\^{e}ncia, Tecnologia e Inova\c{c}\~{a}o 
(Brazil) and Ministerio de Ciencia, Tecnolog\'{i}a e Innovaci\'{o}n Productiva (Argentina).}}

%   \subtitle{I. A young alpha-enriched and lithium-rich giant in the Kepler field}

   \author{E. Jofr\'{e}\inst{1,2}, R. Petrucci\inst{1,2}, L. Garc\'ia\inst{1} and M. G\'{o}mez\inst{1,2} 
           }

   \institute{Observatorio Astron\'{o}mico de C\'{o}rdoba (OAC), Laprida 854, X5000BGR, C\'ordoba, Argentina \\
              \email{emiliano@oac.uncor.edu}
         \and
            Consejo Nacional de Investigaciones Cient\'{i}ficas y T\'{e}cnicas (CONICET), Argentina  
             }

   \date{Received September 10, 2015; accepted October 23, 2015}

% \abstract{}{}{}{}{} 
% 5 {} token are mandatory
 
  \abstract
{We report the discovery of a new exceptional young lithium-rich giant, KIC 9821622, in the \textit{Kepler} field that exhibits an unusually large enhancement of $\alpha$, Fe-peak, and \textit{r}-process elements. From high-resolution spectra obtained with GRACES at Gemini North, we derived fundamental parameters and detailed chemical abundances of 23 elements from equivalent widths and synthesis analysis. By combining atmospheric stellar parameters with available asteroseismic data, we obtained the stellar mass, radius, and age.   
The data analysis reveals that KIC 9821622 is a Li-rich (A(Li)$_{NLTE}$ = 1.80 $\pm$ 0.2) intermediate-mass giant star ($M$ = 1.64 $M_{\sun}$) located at the red giant branch near the luminosity bump. We find unexpectedly elevated abundances of Fe-peak and \textit{r}-process elements. In addition, as previously reported, we find that this is a young star (2.37 Gyr) with unusually high abundances of $\alpha$-elements ([$\alpha$/Fe] = 0.31). The evolutionary status of KIC 9821622 suggests that its Li-rich nature is the result of internal fresh Li that is synthesized through the Cameron-Fowler mechanism near the luminosity bump. However, its peculiar enhancement of $\alpha$, Fe-peak, and \textit{r}-process elements opens the possibility of external contamination by material enriched by a supernova explosion. Although it is less likely, planet accretion cannot be ruled out.}


\authorrunning{Jofr\'{e} et al.}
\titlerunning{KIC 9821622: A interesting lithium-rich giant in the \textit{Kepler} field}

   \maketitle
%
%________________________________________________________________

\section{Introduction}
Lithium (Li) is easily destroyed in the stellar interiors at relatively low temperatures (T $\approx$ 2.5 $\times$ 10$^{6}$ K) by proton-capture reactions. Studying the lithium abundance (A(Li))\footnote{A(Li) = $\log [n(\text{Li})/n(\text{H})]$ + 12} in stellar photospheres therefore is an important tool for understanding stellar evolution. During the main-sequence (MS) phase Li is preserved only in the outermost surface layers. As soon as the star evolves into the red giant branch (RGB) phase, the surviving surface Li is greatly diluted during the first dredge-up (FDU), when the deepening of the convective zone brings high-temperature material to the stellar surface. According to standard models, red giant stars are therefore expected to present low lithium abundances. A star leaving the MS with a meteoritic Li abundance (A(Li) $\simeq$ 3.3 dex) is predicted to have A(Li) $\lesssim$ 1.5 dex on the RGB \citep{Iben1967a}. In fact, observations show that most of the K giants exhibit lower Li abundances \citep{Brown1989, Mallik1999}, suggesting significant Li depletion during the MS or the pre-MS that is due to extra-mixing processes \citep[e.g.,][and references therein]{Charbonnel2000, CantoMartins2011}.

However, results from different surveys reveal that about 1-2\%  of all observed giants have A(Li) $\geq$ 1.5 \citep[][]{Brown1989, Kumar2011, Monaco2011, Lebzelter2012, Liu2014, Adamow2014}. According to the general consensus, these giants are termed Li-rich, and those even more rare that exceed the meteoritic value are called super Li-rich. Several scenarios have been proposed to explain these unexpected objects. The most favored scenario involves a fresh Li production phase following the FDU dilution \citep[see, e.g.,][]{Sackmann1999} through the conversion of $^{3}$He via $^{7}$Be to $^{7}$Li by the Cameron-Fowler mechanism \citep[][CF hereafter]{Cameron1971}. Although the location of the onset of this process is uncertain, it is thought that it can take place in intermediate-mass stars (4--8 M$_{\sun}$) during their evolution on the asymptotic giant branch (AGB) or at the luminosity bump (LB) along the first ascent to the RGB for low-mass stars \citep{Charbonnel2000, Gratton2004}. In agreement with this picture, observational results show that Li-rich giants tend to clump around the LB for low-mass stars and at the red clump for intermediate-mass stars \citep{Charbonnel2000, Reddy2005, Kumar2011}. 

Nevertheless, other recent surveys reveal that Li-rich giants can be found all along the RGB \citep[e.g.,][]{Monaco2011, Lebzelter2012}. As the star continues its evolution along the RGB, the freshly synthesized lithium at the bump begins to be destroyed again when it is mixed with high-temperature material. Thus, Li-rich giants observed near the tip of the RGB may challenge the aforementioned scenario. External pollution caused by the accretion of planets or brown dwarfs \citep{Alexander1967, Siess1999, Denissenkov2004, Adamow2012}, material from evolved AGB companions \citep{Sackmann1999} or produced in core-collapse type II supernova (SNII) explosions \citep{Woosley1995} have been proposed as alternative explanations for the existence of Li-rich giants. In these cases, enrichment patterns in other elements are also expected. 

To advance in our understanding of these rare objects, it is essential not only to continue the search of Li-rich giants, but also to derive their chemical abundances and to unambiguously establish their evolutionary status. In this letter, we report the first Li-rich giant discovered with the new high-resolution spectrograph at Gemini North. This star has also recently been reported to be a young $\alpha$-enriched giant in the solar neighborhood \citep{Martig2015}. 
%__________________________________________________________________

\section{Observations and data reduction}
Along with nine other objects, \object{KIC 9821622} was observed as one of the first science targets taken for an on-sky test of the Gemini Remote Access to CFHT ESPaDOnS Spectrograph \citep[GRACES;][]{Chene2014}. Through a 270 m fiber optics, GRACES integrates the large collecting area of the Gemini North telescope (8.1 m) with the high resolving power and efficiency of the bench-mounted ESPaDOnS spectrograph \citep{Donati2003} at the Canada-France-Hawaii Telescope (CFHT), achieving a maximum resolution power of R$\sim$67500 between 400 and 1000 nm. KIC 9821622 was observed on 21 July 2015 in the one-fiber mode (object-only) with an exposure time of 3 $\times$ 180 s. The spectra were reduced with the OPERA\footnote{OPERA (Open source Pipeline for ESPaDOnS Reduction and Analysis) is available at http://www.cfht.hawaii.edu/en/projects/opera/} software \citep[][Malo et al., in Prep]{Martioli2012}. After correcting for radial-velocity shifts, these data were coadded to obtain a final spectrum with a signal-to-noise ratio of $\sim$150 around 6700 \AA.

\section{Analysis}
The basic properties of KIC 9821622 including atmospheric, asteroseismic, and stellar parameters, are listed in Table \ref{table1}, while the detailed chemical abundances are reported in Table \ref{table2}.
 
\subsection{Fundamental parameters and projected stellar rotation}
Precise spectroscopic fundamental parameters --effective temperature ($T_{\mathrm{eff}}$), surface gravity ($\log g_{spec}$), metallicity ([Fe/H]) and microturbulent velocity ($v_{t}$)-- were derived following the procedure described in \citet{Jofre2015}. Briefly, the atmospheric parameters are calculated from the equivalent widths (EWs) of iron lines (\ion{Fe}{I} and \ion{Fe}{II}) by imposing excitation and ionization equilibrium and the independence between abundances and EWs through the FUNDPAR program \citep{Saffe2011}, which uses the MOOG code \citep{Sneden1973} and 1D LTE ATLAS9 plane-parallel model atmospheres \citep{Kurucz1993}. The EWs of 54 \ion{Fe}{I} and 8 \ion{Fe}{II} lines were automatically computed through the upgraded version of the ARES code \citep{Sousa2015}. In addition, we derived the projected rotational velocity ($v\sin i$) by spectral synthesis of six relatively isolated iron lines following the procedure of \citet{Carlberg2012}.

\subsection{Detailed chemical abundances}
\begin{figure}
   \centering
   \includegraphics[width=.42\textwidth]{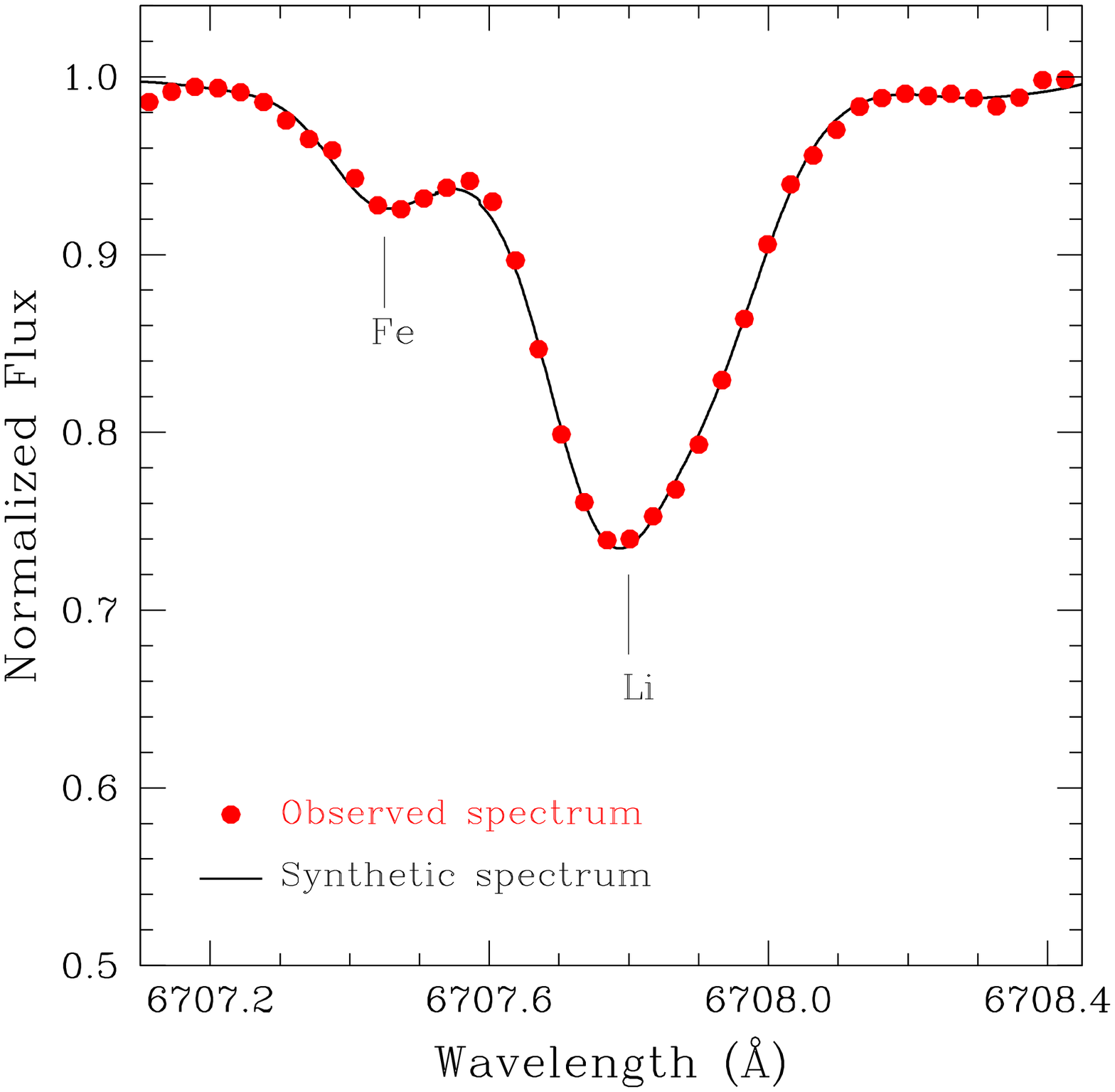}
   \caption{Best fit obtained between the synthetic and the observed GRACES spectra of KIC 9821622 around the 6707.8 {\AA} lithium line.}
              \label{Li-synth}%
    \end{figure}

The abundances of Li, C, N, O, and the carbon $^{12}$C/$^{13}$C isotopic ratio were derived by fitting synthetic spectra to the data using the MOOG code (\textit{synth} driver). For Li we analyzed the \ion{Li}{I} feature at $\lambda$6707.8 {\AA} adopting the line list of \citet{Carlberg2012}. The best fit shown in Fig. \ref{Li-synth} corresponds to an LTE abundance of A(Li) = 1.49 dex, which results in an abundance of A(Li)$_{NLTE}$= 1.65 dex using the NLTE corrections of \citet{Lind2009}. We also measured the weaker but still detected subordinate 6103.6 {\AA} line, using a line list from VALD line database \citep{Kupka1999}. The best fit is obtained for A(Li) = 1.80 dex, which after NLTE corrections results in A(Li)$_{NLTE}$= 1.94 dex. An abundance of A(Li) = 1.80 $\pm$ 0.2  is the mean of NLTE Li abundances from the two Li lines at 6104 and 6708 {\AA}. All these values would classify KIC 9821622 as a Li-rich giant according to the most common criterion ((A(Li) $\geq$ 1.5), although we should mention that our target would be in the limit of Li-normal giants (A(Li) < 1.80) according to other authors \citep[e.g.,][]{Ruchti2011, Liu2014}. 

The oxygen abundance was derived from the analysis of the $\lambda$7771-5 {\AA} infrared (IR) triplet, and the $\lambda$6300, 6363 {\AA} lines. For the abundance obtained from the IR triplet we applied the empirical NLTE corrections by \citet{Afsar2012}. Carbon was derived by fitting the features at 8335 and 9061.43 {\AA} \citep{Siqueira-Mello2015}, while the nitrogen abundance was determined based on CN bands near 8300 {\AA}. The $^{12}$C/$^{13}$C ratio was derived by synthesizing the $\lambda$8002-8004.65 {\AA} region. % using the line list of \citet{Carlberg2012}. 

Additionally, we measured abundances of two odd-Z (Na, Al), four $\alpha$ (Mg, Si, Ca, Ti), seven Fe-peak (Sc, V, Cr, Mn, Co, Ni, Zn), and six neutron capture elements (\ion{Y}{II}, \ion{Ba}{II}, \ion{La}{II}, \ion{Pr}{II}, \ion{Nd}{II}, \ion{Eu}{II}), following the method outlined in \citet{Jofre2015}. Basically, abundances are derived from the EWs through LTE analysis with the MOOG code (\textit{abfind} driver) using the previously calculated model atmosphere. The line list and atomic parameters for these elements were compiled from \citet{Takeda2008}, \citet{Jacobson2013}, \citet{Adibekyan2015}, and \citet{Jofre2015}, and the EWs were measured by Gaussian fitting using the IRAF\footnote{IRAF is distributed by the National Optical Astronomy Observatories, which are operated by the Association of Universities for Research in Astronomy, Inc., under cooperative agreement with the National Science Foundation.}\texttt{splot} task. Our derived fundamental parameters and chemical abundances (odd-Z, $\alpha$, and Fe-peak elements) agree with those obtained from H-band APOGEE spectra \citep{Pinsonneault2014, Holtzman2015}.  

 \begin{figure}
   \centering
   \includegraphics[width=.42\textwidth]{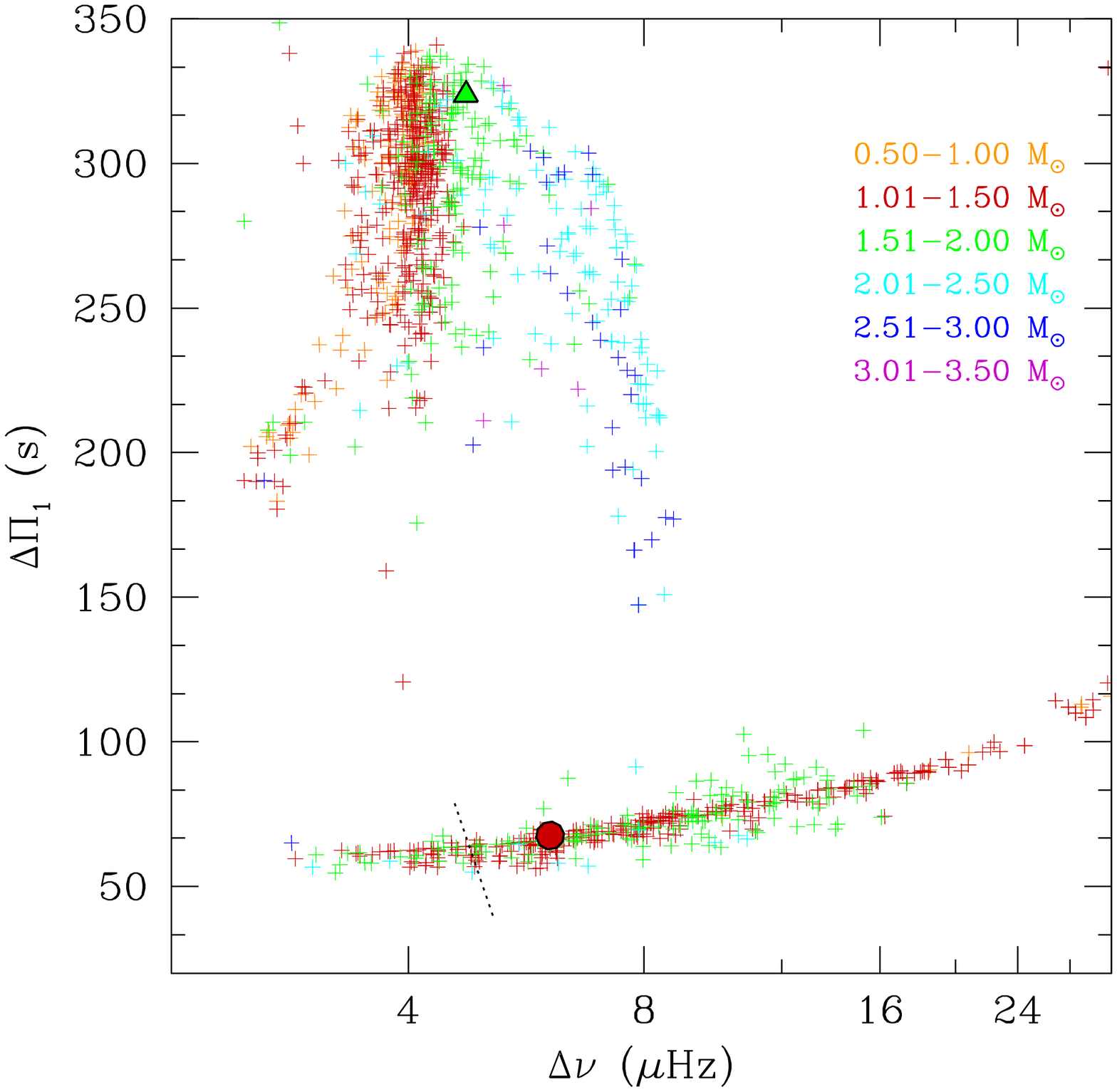}
      \includegraphics[width=.42\textwidth]{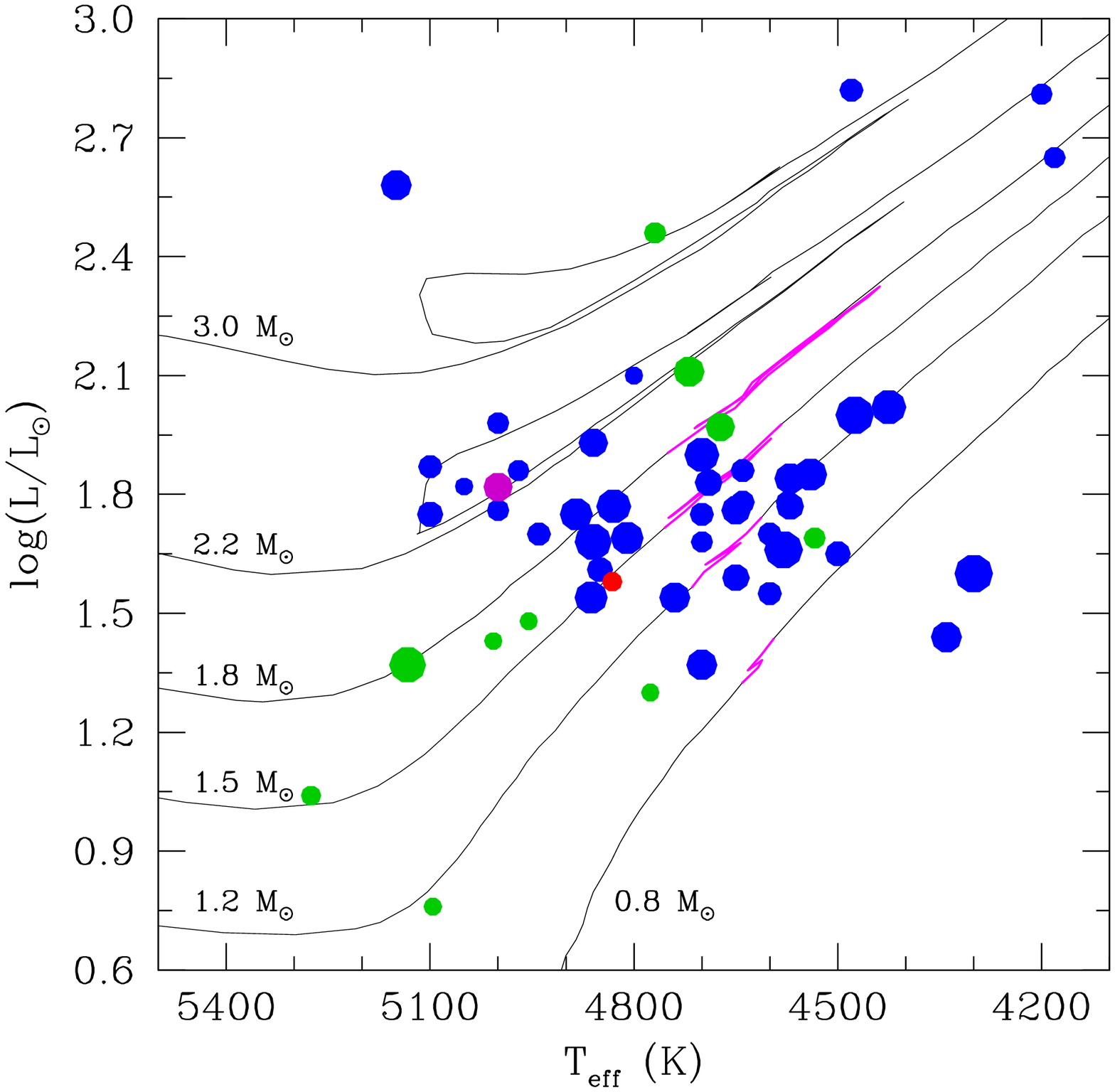}
   \caption{\textit{Top:} $\Delta \Pi_{1}$ vs. $\Delta \nu$ for the \textit{Kepler} sample analyzed by \citet{Mosser2012a} and \citet{Mosser2014}. The location of KIC 9821622 is shown with a filled circle and that of KIC 5000307 with a filled triangle. The stellar mass is indicated by color code and the dotted line represents the approximate position of the LB given by \citet{Mosser2014} for a 1.4 M$_{\sun}$ star. \textit{Bottom:} Position of KIC 9821622 (red) in the standard HR diagram along with other Li-rich giants reported by \citet[][purple]{SilvaAguirre2014}, \citet[][blue]{Kumar2011}, \citet[][]{Adamow2014}, and \citet[][green]{Adamow2015}. Symbol sizes are proportional to the A(Li). Evolutionary tracks computed for [Fe/H] = -0.4 dex from \citet{Bertelli2008} are also depicted. The location of the LB is indicated in magenta.}
              \label{hr}%
    \end{figure}

\subsection{Evolutionary status and stellar parameters}
Asteroseismic analysis from \textit{Kepler} high-precision photometry has revealed that giant stars exhibit not only pressure modes (p-modes) but also gravity modes (g-modes) oscillations. These modes are equally spaced in period, and their average period spacing ($\Delta \Pi_{1}$) gives information about the central cores of red giants, while the pure p-modes are equally spaced in frequency with a separation of $\Delta \nu$ and provide insight about the outer convective envelope. By examining the variation of $\Delta \Pi_{1}$ as a function of  $\Delta \nu$, it is therefore possible to unambiguously distinguish between different late evolutionary states \citep{Bedding2011, Mosser2012a, Stello2013}. The upper panel of Fig. \ref{hr} shows the seismic $\Delta\nu$-$\Delta\Pi_{1}$ HR diagram of the \textit{Kepler} sample analyzed by \citet{Mosser2012a} and \citet{Mosser2014}. Here, it can be seen that our target (filled circle) is unequivocally located at the RGB region ($\Delta\Pi_{1}$ $\leq$ 100 s), indicating that this Li-rich giant is only burning hydrogen in a shell. In the clump region ($\Delta\Pi_{1}$ $\geq$ 150 s) we also show the location of the first confirmed Li-rich core-helium burning giant in the \textit{Kepler} field (KIC 5000307, filled triangle) reported by \citet{SilvaAguirre2014}. 
 
From the asteroseismic parameters $\Delta \nu$ and $\nu_{max}$ (frequency of maximum oscillation power) along with the derived $T_{eff}$ and [Fe/H], we computed the stellar mass, radius, and age of KIC 9821622 through a Bayesian estimation method using the PARAM\footnote{Version 1.3: http://stev.oapd.inaf.it/cgi-bin/param\_1.3} code \citep{daSilva2006}. %with the new PARSEC isochrones \citep{Bressan2012}.}    

\section{Discussion}
The bottom panel of Fig. \ref{hr} shows the position of our target (red) in the HR diagram along with the evolutionary tracks from \citet{Bertelli2008}. We also indicate the Li-rich giants (A(Li)$_{NLTE}$ $\geq$ 1.5 dex) reported by \citet[][green]{Adamow2014, Adamow2015}, \citet[][purple]{SilvaAguirre2014}, and those identified and collected by \citet[][blue]{Kumar2011}. From this figure and the asteroseismic HR diagram (upper panel), it is clear that KIC 9821622 is slightly below the RGB LB phase along with other Li-rich giants. As suggested by \citet{Kumar2011}, these pre-bump giants are very likely LB stars. In this case, as most Li-rich giants on this location, the Li-enhancement observed for our target would be consistent with a scenario where fresh Li is synthesized through the CF mechanism. Here, the extra-mixing necessary to feed $^{3}$He to the hotter stellar interior is associated with the removal of the mean molecular weight discontinuity arising from the FDU \citep{Charbonnel2000}. 

Another possibility, suggested by \citet{Denissenkov2012}, is that stars at this location in the HR diagram are objects that have already reached the LB and are now making extended zigzags toward lower luminosities. This would be the result of their fast internal rotation and associated turbulent mixing that also would produce an enhancement in the Li content. Predictions of this model include lower isotopic ratios than the standard post-FDU value ($^{12}$C/$^{13}$C $\approx$ 25). The internal rotational splitting \citep[$\delta \nu _{rot}$ $\gtrsim$ 200 nHz;][]{Mosser2012b} and measured carbon isotopic ratio ($^{12}$C/$^{13}$C = 18) of KIC 9821622 might support this alternative.  
 
In an external pollution scenario we should expect to detect other chemical anomalies in addition to the enhancement of lithium. An evolved AGB binary companion could be producing Li through the CF mechanism and be polluting the surface of KIC 9821622. A transfer of Li would be accompanied by an enrichment in the abundances of C, N, O, and \textit{s}-process elements such as Ba, Y, and La. The derived abundances of all these elements, except for O, which is particularly high, are consistent with the abundances of giants with similar metallicities in the solar neighborhood \citep{Luck2007, Takeda2008}. Furthermore, there is no signature of binarity in the \textit{Kepler} light curves or in the spectra. However, a systematic radial velocity follow-up of this target is necessary to firmly rule-out the existence of a binary companion. 

Otherwise, Li enhancement in giant stars can be the result of the engulfment of a brown dwarf or planet \citep{Siess1999}. Using Eq. 2 of \citet{Siess1999}, we can estimate the mass of the accreted body to reproduce the observed lithium excess. Adopting a convective envelope of 0.7 $M_{\sun}$ and a meteoritic lithium abundance for the hypothetically accreted body, we estimated a mass of $\sim$0.23 $M_{Jup}$, which is a realistic size for a planet. It has been pointed out that an engulfment episode would also increase the rotational velocity of the host star \citep{Siess1999, Carlberg2012}. The $v\sin i$ of KIC 9821622 (= 1.01 km/s) is typical for evolved stars. However, given the estimated mass of the potential accreted planet, this low rotation would not necessarily contradict the engulfment scenario. Moreover, \citet{Siess1999} suggest that the accretion of a planet would produce the formation of a shell of ejected material that can be detected as IR excess, although this might be true independently of the mechanism that produces the lithium enhancement \citep[see, e.g.,][]{delaReza1996, delaReza1997}. The analysis of the spectral energy distribution \footnote{Constructed from photometry of KIS \citep{Greiss2012}, 2MASS \citep{Skrutskie2006}, and WISE \citep{Cutri2012}, the star's distance from \citet{Rodrigues2014} and assuming a blackbody curve to fit the stellar continuum.} for KIC 9821622, shows a hint of IR excess at 22 $\mu$m. Using the test given by \citet{Rebull2015}, which makes use of the [3.4]--[22] WISE data, we confirm a small IR excess at 22 $\mu$m. Observations at longer wavelengths (mid-IR and submm) would be highly valuable for constraining the parameters of the possible circumstellar shell or ring of dust. 

Mass-loss episodes can also cause blueshifted asymmetric H$_{\alpha}$ profiles or additional absorption components in the wings of the Na doublet lines \citep[see][and references therein]{Monaco2011}. None of these effects is observed in our spectra, although Na lines are slightly asymmetric. Furthermore, the analysis of the IR \ion{Ca}{II} triplet (8498--8662 {\AA}) does not show the reversal typical for active stars. Additional constraints for the planet accretion scenario may come from the detection of beryllium enhancement analyzing the line at $\lambda$3131 {\AA}, which unfortunately is beyond our spectral coverage. 

Finally, external pollution of material produced by SNII explosions should also enhance the abundances of $\alpha$, Fe-peak and \textit{r}-process elements. Consistent with this scenario, we find unusually high values of O, Si, Ca, Mg, and Ti giving a [$\alpha$/Fe] = 0.31, which is particularly high for the stellar age of 2.37 Gyr \citep[see also][]{Martig2015, Chiappini2015}. Moreover, the abundances of several Fe-peak elements (V, Cr, Sc, Ni) and \textit{r}-process elements (Eu, Nd) seem to depart significantly from the trend found for nearby giants \citep{Luck2007, Takeda2008}, providing support to this scenario. On the other hand, according to \citet{Chiappini2015}, the existence of $\alpha$-rich objects is not explained by standard chemical evolution models of the Milky Way, and tentatively suggest that these stars could be formed near the end of the Galactic bar. However, \citet{Martig2015} were unable to confirm different birth locations for the $\alpha$-rich young stars compared to the other $\alpha$-rich population from their \textit{Kepler} sample. Another possibility suggested by Chiappini et al. is that these objects formed from a recent gas accretion event. If this were the case, the Li excess of this giant would be the product of some of the aforementioned mechanisms. 

KIC 9821622 is certainly a unique and interesting object that deserves further scrutiny to reveal the real mechanism behind the observed anomalous abundances. In this sense, high-resolution chemical analysis of more of these young $\alpha$-rich giants might help to understand their origin (Jofr\'e et al., in prep.).

\begin{acknowledgements}
      We thank Lison Malo and Eder Martioli for reducing the spectra used in this work. E. J. and R. P. acknowledge the financial support from CONICET in the forms of Post-Doctoral Fellowships. We also thank the anonymous referee for helpful and valuable comments and suggestions. 
\end{acknowledgements}


\bibliographystyle{aa}
\bibliography{ref.bib} 

\Online

 \onltab{
   \begin{table*}
   \tiny
      \caption[]{Fundamental properties of KIC 9821622.}
         \label{table1}
     \centering
         \begin{tabular}{l l l }
            \hline\hline
              Parameter &  Value & Reference \\
            \hline
RA (J2000)  &   19:08:36.1   &\\
Dec (J2000) &   +46:41:41.2   &\\
2MASS Id & J19083615+4641212 & \\
K$_{2MASS}$ [mag] & 9.83 & \citet{Skrutskie2006}\\
Distance [pc] & 1615   & \citet{Rodrigues2014} \\
\hline
$T_{\mathrm{eff}}$ [K] & 4725 $\pm$ 20 & This work\\
$\log g_{spec}$ & 2.73 $\pm$ 0.09 & This work\\
$v_{t}$ [km $s^{-1}$]& 1.12 $\pm$ 0.04 & This work \\
\text{[Fe/H]} & -0.49 $\pm$ 0.03 & This work\\
v$\sin i$ [km $s^{-1}$] & 1.01 $\pm$ 0.77 & This work \\
\hline
$\nu_{max}$ [$\mu$Hz] & 63.72 $\pm$ 1.49 & \citet{Mosser2014}\\
$\Delta\nu$ [$\mu$Hz] & 6.07 $\pm$ 0.05  & \citet{Mosser2014}\\
$\Delta\Pi$ [s] & 67.6 $\pm$ 1.30  & \citet{Mosser2014} \\
Mass [$M_{\odot}$] & 1.64 $\pm$ 0.06 & This work\\
Radius [$R_{\odot}$] & 9.33 $\pm$ 0.20 & This work\\
Age [Gyr] & 2.37 $\pm$ 0.58 & This work\\
$\log g_{seis}$ & 2.71 $\pm$ 0.01 & This work \\
$\log (L/L_{\odot})$ & 1.59 $\pm$ 0.10 & This work\\
            \hline
         \end{tabular}     
        \end{table*} 
        
}

 \onltab{
   \begin{table*}
   \tiny
      \caption[]{Derived chemical abundances of KIC 9821622.}
         \label{table2}
     \centering
     
         \begin{tabular}{l c c c c}
            \hline\hline
Element	&	A(X) $\pm$ $\sigma$\tablefootmark{a}			&	\text{[X/H]}	&	\text{[X/Fe]}	&	A(X)$_{\sun}$ \tablefootmark{b}	\\
\hline
Li (6708 \AA)	&	1.49 $\pm$ 0.02			&		&		&		\\
Li (6708 \AA)$_{NLTE}$	&	1.65			&		&		&		\\
Li (6104 \AA	)&	1.80 $\pm$ 0.04			&		&		&		\\
Li (6104 \AA)$_{NLTE}$	&	1.94			&		&		&		\\
C	&	8.15	$\pm$	0.09	&	-0.28	&	0.13	&	8.43	\\
N	&	7.33	$\pm$	0.01	&	-0.50	&	-0.09	&	7.83	\\
O (7771-5 \AA)	&	9.14	$\pm$	0.03	&	0.45	&	0.86	&	8.69	\\
O (7771-5 \AA)$_{NLTE}$	&	8.79	$\pm$	0.04	&	0.10	&	0.51	&	8.69	\\
O (6300/6363 \AA)	&	8.74	$\pm$	0.02	&	0.05	&	0.46	&	8.69	\\
$^{12}$C/$^{13}$C & 18 $\pm$ 0.70 & & & \\
\hline
Na	&	6.06	$\pm$	0.01	&	-0.18	&	0.23	&	6.24	\\
Mg	&	7.47	$\pm$	0.08	&	-0.13	&	0.28	&	7.60	\\
Al	&	6.39	$\pm$	0.04	&	-0.06	&	0.35	&	6.45	\\
Si	&	7.40	$\pm$	0.06	&	-0.11	&	0.30	&	7.51	\\
Ca	&	6.13	$\pm$	0.04	&	-0.21	&	0.20	&	6.34	\\
Sc	&	2.84	$\pm$	0.03	&	-0.31	&	0.10	&	3.15	\\
Ti	&	4.82	$\pm$	0.04	&	-0.13	&	0.28	&	4.95	\\
V	&	3.70	$\pm$	0.03	&	-0.23	&	0.18	&	3.93	\\
Cr	&	5.23	$\pm$	0.04	&	-0.41	&	0.00	&	5.64	\\
Mn	&	4.89	$\pm$	0.03	&	-0.54	&	-0.13	&	5.43	\\
Co	&	4.62	$\pm$	0.04	&	-0.37	&	0.04	&	4.99	\\
Ni	&	5.84	$\pm$	0.04	&	-0.38	&	0.03	&	6.22	\\
Zn	&	4.51	$\pm$	0.01	&	-0.05	&	0.36	&	4.56	\\
Y II	&	1.77	$\pm$	0.06	&	-0.44	&	-0.03	&	2.21	\\
Ba II	&	1.63	$\pm$	0.06	&	-0.55	&	-0.14	&	2.18	\\
La II	&	0.85	 		&	-0.28	&	0.13	&	1.13	\\
Pr II	&	0.04		 	&	-0.67	&	-0.26	&	0.71	\\
Nd II	&	1.09	$\pm$	0.17	&	-0.36	&	0.05	&	1.45	\\
Eu II	&	0.63	$\pm$	0.12	&	0.11	&	0.52	&	0.52	\\
\hline
         \end{tabular}
\tablefoot{
\tablefoottext{a}{For Li, N, and $^{12}$C/$^{13}$C the $\sigma$ value represents the uncertainty of the spectral synthesis fit, while for the remaining elements $\sigma$ is the standard deviation of the mean abundance obtained from all the measured lines.}
\tablefoottext{b}{Adopted solar values from \citet{Asplund2009}. }

}
     
        \end{table*} 
}

\listofobjects
     
\end{document}